\def\ptitle{\tiny Generalized spiked harmonic oscillator $\dots$}
\font\tr=cmr12                          
\font\bf=cmbx12                         
\font\sl=cmsl12                         
\font\it=cmti12                         
\font\trbig=cmbx12 scaled 1500          
\font\tiny=cmr10                        
\output={\shipout\vbox{\makeheadline
                                      \ifnum\the\pageno>1 {\hrule}  \fi 
                                      {\pagebody}   
                                      \makefootline}
                   \advancepageno}

\headline{\noindent {\ifnum\the\pageno>1 
                                   {\tiny \ptitle\hfil
page~\the\pageno}\fi}}
\footline{}

\tr 
\def\bra{{\rm <}}    
\def\ket{{\rm >}}    
\def\nl{\hfil\break\noindent}  
\def\ni{\noindent}             
\def\np{\hfil\vfil\break}
\def\htab#1#2{{\hskip #1 in #2}}
\baselineskip 15 true pt  
\parskip=0pt plus 5pt 
\parindent 0.25in
\hsize 6.0 true in 
\hoffset 0.25 true in 
\emergencystretch=0.6 in                 
\vfuzz 0.4 in                            
\hfuzz  0.4 in                           
\vglue 0.1true in
\mathsurround=2pt                        
\topskip=24pt                            
\newcount\zz  \zz=0  
\newcount\q   
\newcount\qq    \qq=0  

\def\pref #1#2#3#4#5{\frenchspacing \global \advance \q by 1     
    \edef#1{\the\q}
       {\ifnum \zz=1 { %
         \item{[\the\q]} 
         {#2} {\bf #3},{ #4.}{~#5}\medskip} \fi}}

\def\bref #1#2#3#4#5{\frenchspacing \global \advance \q by 1     
    \edef#1{\the\q}
    {\ifnum \zz=1 { %
       \item{[\the\q]} 
       {#2}, {\it #3} {(#4).}{~#5}\medskip} \fi}}

\def\gref #1#2{\frenchspacing \global \advance \q by 1  
    \edef#1{\the\q}
    {\ifnum \zz=1 { %
       \item{[\the\q]} 
       {#2.}\medskip} \fi}}

\def\oaint{{\oint\limits_{{\cal C}}\!\!\!\!\;{}_{\hat{}}}}

 \def\sref #1{~[#1]}

\def\references#1{\zz=#1
   \parskip=2pt plus 1pt   
   {\ifnum \zz=1 {\noindent \bf References \medskip} \fi} \q=\qq

\pref{\detw}{L. C. Detwiler and J. R. Klauder, Phys. Rev. D}{11}{1436 (1975)}{}

\pref{\harr}{E. M. Harrell, Ann. Phys. }{105}{ 379 (1977)}{}

\pref{\agua}{V. C. Aguilera-Navarro, G.A. Est\'evez, and R.
Guardiola, J. Math. Phys.}{31}{99 (1990)}{}

\pref{\agub}{V. C. Aguilera-Navarro and R. Guardiola, J. Math.
Phys.}{32}{2135 (1991)}{}

\pref{\sol}{Solano-Torres, G. A. Est\'eves, F. M. Fern\'andez,
and G. C. Groenenboom, J. Phys. A: Math. Gen.}{25}{3427 (1992)}{}

\pref{\flyn}{M. F. Flynn, R. Guardiola, and M. Znojil, Czech. J.
Phys.}{41}{1019 (1993)}{}

\pref{\znoa}{M. Znojil, Proc. Int. Conf. on Hadron Structure}{91}{Proc. CSFR vol.1, 1 (1992)}{}

\pref{\znob}{M. Znojil, J. Math. Phys.}{34}{4914 (1993)}{}

\pref{\\nag}{N. Nag and R. Roychoudhury, Czech. J. Phys.}{46}{343 (1996)}{} 

\pref{\znoc}{M. Znojil and R. Roychoudhury, Czech. J. Phys.}{48}{1 (1998)}{}

\pref{\esta}{E. S. Est\'evez-Bret\'on and G. A. Est\'evez-Bret\'on,  J. Math. Phys.}{34}{437 (1993)}{}

\pref{\hala}{R. Hall, N. Saad and A. von Keviczky, J. Math. Phys.}{39}{6345-51 (1998)}{}
 
\pref{\halb}{R. Hall and N. Saad, J. Phys. A: Math. Gen.}{33}{569 (2000)}{}

\pref{\halc}{R. Hall and N. Saad, J. Phys. A: Math. Gen.}{32}{133 (1999)}{}

\gref{\muod}{O. Mustafa, M. Odeh, e-print quant-ph/006004}

\gref{\jski}{J. Skibi\'nski, e-print quant-ph/0007059}

\pref{\klau}{J. R. Klauder, Acta Phys. Austriaca Suppl.}{11}{
341 (1973)}{}

\pref{\simo}{B. Simon, J. Functional Anal.}{14}{295 (1973)}{}

\pref{\defa}{B. DeFacio and C. L. Hammer, J. Math. Phys.}{15}{1071 (1974)}{}

\bref{\gold}{I. I. Gol'dman and D. V. Krivchenkov}{Problems in Quantum mechanics}{Pergamon, London, 1961}{}

\bref{\land}{L. D. Landau and E. M. Lifshitz}{Quantum mechanics: non-relativistic theory}{Pergamon, London, 1981}{}

\pref{\eks}{H. Ezawa, J. R. Klauder, and L. A. Shepp, J. Math. Phys.}{16}{783 (1975)}{}

\pref{\BS}{H. Behnke and F. Sommer,}{\it Theorie der Analytischen Funktionen einer Komplexen Ver\"anderlichen} {Springer-Verlag, Berlin, 1976}{Eq.(38), page 228; Eq.(41), page 229; Eq.(43), page 330.}

\bref{\slat}{L. J. Slater}{Generalized Hypergeometric
Functions}{University Press, Cambridge, 1966}{}

\bref{\sla}{L. J. Slater}{Confluent Hypergeometric Functions}{University Press, Cambridge, 1960}{}

\bref{\Erd}{A. Erd\'elyi}{Higher Trancendental Function, Vol. II}{Bateman Project, McGraw-Hill, 1953}{}

\bref{\ndlt}{Nota di Letterio Toscanp}{Boll. Un. Mat. Ital.}{3}{398 (1949)}{}

\bref{\jowe}{J. Weidmann}{\it Linear Operators in Hilbert Space}{New York, Springer-Verlag, 1980} {Appendix A, Lebesgue Integration, Theorem A8, pp369-370.}
}

 \references{0}    

\htab{3.5}{CUQM-83}

\htab{3.5}{math-ph/0101006}

\htab{3.5}{January 2001}
\vskip 0.5 true in
\centerline{\bf\trbig Generalized Spiked Harmonic Oscillator}
\medskip
\vskip 0.25 true in
\centerline{Richard L. Hall, Nasser Saad, and Attila B. von Keviczky}
\medskip
{\leftskip=0pt plus 1fil
\rightskip=0pt plus 1fil\parfillskip=0pt\baselineskip 18 true pt
\obeylines
\baselineskip 10 pt
Department of Mathematics and Statistics, Concordia University,
1455 de Maisonneuve Boulevard West, Montr\'eal, 
Qu\'ebec, Canada H3G 1M8.\par}

\vskip 0.5 true in
\baselineskip 15 pt
\centerline{\bf Abstract}\medskip
A variational and perturbative treatment is provided for a family of generalized spiked 
harmonic oscillator Hamiltonians $H=-{d^2\over dx^2}+B x^2+{A\over x^2}+{\lambda\over x^\alpha}$,where $B>0,\ A\geq 0$, and $\alpha$ and $\lambda$ denote two real positive parameters. The method makes use of the function space spanned by the solutions $|n\ket$ of Schr\"odinger's equation for the potential $V(x)= B x^2+{A\over x^2}$. Compact closed-form expressions are obtained for the matrix elements $\bra m|H|n\ket,$ and a first-order perturbation series is derived for the wave function.  The results are given in terms of generalized hypergeometric functions. It is proved that the series for the wave function is absolutely convergent for $\alpha \le 2.$ 


\vfil\eject

\ni{\bf 1. Introduction}\medskip
\ni The term ``supersingular potential'' was introduced by Detwiler and Klauder\sref{\detw} to describe the situation where every matrix element of a perturbation with respect to the unperturbed state is infinite. An important class of problems with this feature is the family of spiked harmonic oscillators with Hamiltonian $H$ given by Harrell\sref{\harr}, where
$$
H=-{d^2\over dx^2}+ x^2 +{\lambda\over x^\alpha}\quad 0\leq x<\infty,\quad \alpha,\lambda > 0,\eqno(1.1)
$$
\nl and $\alpha \ge 3.$  This Hamiltonian has had a variety of applications in atomic, molecular, nuclear and particle physics, where it provides a simple model in which the potential has a repulsive core $\lambda x^{-\alpha}.$  Since the early investigations of Harrell, spiked harmonic-oscillator potentials have become the subject of an intensive study lasting over two decades\sref{\agua-\jski}. Most of this work is concerned with approximations for the energy of the problem in one spatial dimension, because the Klauder phenomenon\sref{\klau-\defa} associated with the Hamiltonian (1.1) fails to occur in higher dimensions.  Meanwhile, no definite results have been obtained so far on a perturbation expansion for the {\it wave function} associated with the Hamiltonian (1.1), although Aguilera-Navarro and Guardiola\sref{\agub} have reported that they met serious convergence difficulties in the case $\alpha=2$ with expressions for the first-order correction to the wave function.  
     
In earlier work\sref{\hala-\halc} we have pointed out the advantages of basing our perturbation analysis on an exactly soluble model which itself has a singular potential term\sref{\hala}. We examined a family of generalized spiked harmonic-oscillator Hamiltonians 
$$
H=H_0+\lambda V=-{d^2\over dx^2}+B x^2+{A\over x^2}+{\lambda\over x^\alpha}, B>0,\ A\geq 0, \eqno(1.2)
$$
defined on the one-dimensional space $(0\leq x <\infty)$ with eigenfunctions satisfying Dirichlet boundary conditions, that is to say, with wave functions vanishing at the boundaries. The singular basis, consisting of the set of exact solutions of $H_0$, serves as a better starting point for the analysis of the singular problem than the Hermite functions used in most earlier work on the Hamiltonian (1.1). In this present paper, we derive compact expressions for the matrix elements developed earlier.  Moreover, we prove convergence of the first order correction to the wave function, using standard Rayleigh-Schr\"odinger perturbation theory for the case $\alpha\leq 2$. Specifically, our main results can be summarized as follows:
\item{1.} The matrix elements of the operator $x^{-\alpha}$, with respect to the exact solutions of the Gol'dman and Krivchenkov Hamiltonian\sref{\gold-\land} $H_0$, are given by the following explicit expressions
$$
x_{mn}^{-\alpha}=(-1)^{n+m}B^{{\alpha\over 4}}{{({\alpha\over 2})_n}\over
(\gamma)_n}{{\Gamma(\gamma-{\alpha\over 2})}\over
\Gamma(\gamma)}\sqrt{{(\gamma)_n(\gamma)_m}\over {n!m!}}{}_3F_{2}(-m,\gamma-{\alpha\over
2},1-{\alpha\over 2};\gamma,1-n-{\alpha\over 2};1),\eqno(1.3)
$$
where $\gamma=1+{1\over 2}\sqrt{1+4A}$, valid for all values of the parameters  $\gamma$ and $\alpha$ such that $\alpha <2\gamma$.  These expressions are also very convenient for direct computations. In particular, these expressions, drawing their importantance from their simplicity, are used in studies of so-called `vestiges'\sref{\eks} of potentials, left after the cofficient $\lambda\downarrow 0.$

\item{2.} The first-order perturbation correction of the wave function of Hamiltonian (1.2), for 
$\alpha =2$, is
$$
\psi_0^{(1)}(x)={B^{-{\gamma\over 4}}
\over 2\sqrt{2}}
{\Gamma(\gamma-1)\over \sqrt{\Gamma(\gamma)}}
x^{\gamma-{1\over 2}}e^{-{\sqrt{B}\over 2}x^2}
[ln(\sqrt{B}x^2)-\psi(\gamma)],\eqno(1.4)
$$ 
where $\psi(\gamma)$ is the digamma function, also known as the logarithmic derivative of the $\Gamma$-function, defined by $\psi(x)={d\over dx}ln \Gamma(x)$.  We note that the coefficient of the exponential term is a logarithmic epression in the variable $x,$ instead of being a polynomial. This fact explains the odd result that the perturbation expansion fails to be regular. In the case of $\alpha<2$, we demonstrate that the first order correction of the wave function is 
$$
\eqalign{
\psi_0^{(1)}(x)&=-{\alpha B^{{\alpha\over 2}-{3\gamma\over 4}}
\over 4\sqrt{2}}
{\Gamma(\gamma-{\alpha\over 2})\over 2\pi\sqrt{\Gamma(\gamma)}}
x^{\gamma-{1\over 2}}e^{-{\sqrt{B}\over 2}x^2}\times\cr
&
\times \int\limits_{-\infty}^{\infty}e^{\sqrt{B} (c+iy)}(c+iy)^{-\gamma}
(1-{x^2\over c+iy}){}_3F_2(1,1,1+{\alpha\over 2};2,2;1-{x^2\over c+iy})dy.}\eqno(1.5)
$$

\noindent The functions  ${}_3F_{2}$ are special cases of the generalized hypergeometric function defined by
$$
{}_pF_{q}(\alpha_1,\alpha_2,\dots,\alpha_p;\beta_1,\beta_2,\dots,\beta_q;z)=\sum\limits_{k=0}^\infty 
{\prod\limits_{i=1}^p(\alpha_i)_k\over  \prod\limits_{j=1}^q(\beta_j)_k}{z^k\over k!},\eqno(1.6)
$$ 
where $p$ and $q$ are non-negative integers and $\beta_j$ ($j=1,2,\dots,q$) cannot be a non-positive integer.  The expression $(a)_k$ denotes the Pochhammer symbol
$$(a_i)_0=1,\quad (a_i)_k=a_i(a_i+1)\dots (a_i+k-1)={\Gamma(a_i+k)\over \Gamma(a_i)},\quad k=1,2,\dots,$$
which we also give in terms of the gamma function $\Gamma(a)$. The function defined by (1.6) includes as special cases the hypergeometric and confluent hypergeometric functions ${}_2F_1$ and ${}_1F_1$ respectively. If the series does not terminate (i.e., none of $\alpha_i$, $i=1,2,\dots,p$, is a negative integer), then the series converges or diverges according as $|z|<1$ or $|z|>1$. For $z=1$ on the other hand, the series is convergent, provided
$
{\sum\limits_{j=1}^q \beta_j-\sum\limits_{i=1}^p \alpha_i}>0.
$ 

\medskip
\noindent{\bf 2. The matrix elements}\medskip
Landau and Lifshitz\sref{\land} exhibited a closed form description of the exact solutions of the one-dimensional Schr\"odinger equation (in units $\hbar=2m=1$)
$$
-\psi^{\prime\prime}_n(x)+(Bx^2+{A\over x^{2}})\psi_n(x)=E_n\psi_n(x),\quad x\in [0,\infty),\eqno(2.1)
$$
where the $\psi_n(x)$ satisfy the Dirichlet boundary condition $\psi_n(0)=0$. The wave functions have the form
$$\psi_n(x)\equiv |n>=(-1)^n\sqrt{{2B^{\gamma\over 2}\Gamma(n+\gamma)}\over n!\Gamma^2(\gamma)}x^{\gamma-{1\over 2}}e^{-{\sqrt{B}\over 2}x^2}{}_1F_1(-n,\gamma,\sqrt{B}x^2),\eqno(2.2)
$$
with exact eigenenergies
$$E_n=2\sqrt{B}(2n+\gamma),\quad n=0,1,2,\dots,\eqno(2.3)$$
for
$\gamma=1+{1\over 2}\sqrt{1+4A}$. Herein, ${}_1F_1$ stands for the confluent hypergeometric function defined in (1.6). We have further introduced the alternating coefficient terms $(-1)^n$ in the definition of $\psi_n(x)$ in order to guarantee a smooth transition, through the identity
$${}_1F_1(-n,{3\over 2},x^2)={(-1)^n\over 2x}{n!\over (2n+1)!}H_{2n+1}(x),\eqno(2.4)$$
to the solutions of the harmonic oscillator problem $A=0$.

Using the integral representation of the confluent hypergeometric function, Hall {\it et al} proved that the matrix elements of the generalized spiked harmonic oscillator for $\alpha<2\gamma$ are
$$
x_{mn}^{-\alpha}=
(-1)^{m+n}
B^{\alpha\over 4}\sqrt{{{\Gamma(\gamma+m)}\over
{n!m!\Gamma(\gamma+n)}}}{{\Gamma(\gamma-{\alpha\over 2})}\over\Gamma(\gamma)}
\sum\limits_{k=0}^m(-1)^k {m\choose k}
{
{\Gamma(k+\gamma-{{\alpha}\over 2})\Gamma({\alpha\over 2}-k+n)}
\over
{\Gamma(k+\gamma)\Gamma({\alpha\over 2}-k)}}.\eqno(2.5)
$$
These expressions can be readily obtained and further simplified by means of the series representation of the confluent hypergeometric function (2.4). Indeed, using (2.2) and (2.4), it immediately follows from
$$T_s=(-1)^s\sqrt{{2B^{\gamma\over 2}\Gamma(s+\gamma)}\over s!\Gamma^2(\gamma)}\eqno(2.6)$$
that
$$<m|x^{-\alpha}|n>=T_nT_m\sum\limits_{k=0}^m\sum_{l=0}^n {{(-m)_k(-n)_l}\over (\gamma)_k(\gamma)_l}{B^{{1\over 2}(k+l)}\over k! l!}\int_0^\infty x^{-\alpha+2\gamma+2k+2l-1}e^{-\sqrt{B}x^2}dx.
$$
By resorting to the integral representation of gamma function, we get for $ -{\alpha\over 2}+\gamma+k+l>0$ that
$$\eqalign{<m|x^{-\alpha}|n>&={1\over 2}T_nT_m\sum\limits_{k=0}^m \sum_{l=0}^n {{(-m)_k(-n)_l}\over (\gamma)_k(\gamma)_l}
{B^{{\alpha\over 4}-{\gamma\over 2}}\over k! l!}\Gamma(-{\alpha\over 2}+\gamma+k+l)\cr
&={1\over 2}T_nT_m B^{{\alpha\over 4}-{\gamma\over 2}}\sum\limits_{k=0}^m
\bigg[\sum_{l=0}^n{{(-n)_l\Gamma(-{\alpha\over 2}+\gamma+k+l)}\over (\gamma)_l\ l!}\bigg]{(-m)_k\over (\gamma)_k}{1\over k!}.}\eqno(2.7)
$$
On the other hand, by use of the definition of the Pochhammer symbols and the series representation of the hypergeometric function (1.6), the finite sum inside the bracket collapses to
$$
\eqalign{\sum_{l=0}^n{{(-n)_l\Gamma(-{\alpha\over 2}+\gamma+k+l)}\over (\gamma)_l\ l!}&=\sum_{l=0}^n{{(-n)_l\ (-{\alpha\over 2}+\gamma+k)_l}\over (\gamma)_l\ l!}\Gamma(-{\alpha\over 2}+\gamma+k)\cr
&=\Gamma(-{\alpha\over 2}+\gamma+k){}_2F_1(-n,-{\alpha\over 2}+\gamma+k;\gamma;1)\cr
&=\Gamma(-{\alpha\over 2}+\gamma+k){({\alpha\over 2}-k)_n\over (\gamma)_n},}\eqno(2.8)
$$
after we invoke Vandermonde's theorem\sref{\slat}. Consequently, we arrive at the matrix elements
$$\eqalign{<m|x^{-\alpha}|n>&={1\over 2}T_nT_m 
B^{{\alpha\over 4}-{\gamma\over 2}}\sum\limits_{k=0}^m
\Gamma(-{\alpha\over 2}+\gamma+k){({\alpha\over 2}-k)_n\over (\gamma)_n}{(-m)_k\over (\gamma)_k}{1\over k!}
\cr
&={1\over 2}T_nT_mB^{{\alpha\over 4}-{\gamma\over 2}}{\Gamma(\gamma)\over (\gamma)_n}
\sum\limits_{k=0}^m(-1)^k{m\choose k}
{\Gamma({\alpha\over 2}+n-k)\Gamma(-{\alpha\over 2}+\gamma+k)\over \Gamma(\gamma+k)\Gamma({\alpha\over 2}-k)},}\eqno(2.9)
$$
wherein we have used the identity ${(-m)_k\over k!}=(-1)^k{m\choose k}$. After writing $T_n$ and $T_m$ for the expressions defined by (2.6), this leads us clearly to the matrix elements (2.5). 
If we now factor out $m!$ from the finite sum appearing in equation (2.9), then we are left with 
$$ \sum\limits_{k=0}^m (-1)^k {1\over k! (m-k)!} {\Gamma(k + \gamma - \alpha /2)
\Gamma(\alpha /2 - k + n)\over \Gamma(k + \gamma)\Gamma(\alpha /2 - k)} = 
\sum\limits_{k=0}^m f(k), \quad f(z) = e^{i \pi z}\times$$
$${\Gamma(z + \gamma - \alpha /2)\Gamma(\alpha /2 - z + n)
\over \Gamma(z + 1)\Gamma(m + 1 - z) \Gamma(z + \gamma)\Gamma(\alpha /2 - z)},\eqno(2.10)$$ 
where $f(z)$ has at most removable singularities at $z = \alpha /2 + n + r$ for $r$ a positive integer such that $\alpha /2 + n + r \in (-1/2, m + 1/2)$. This $f(z)$ is therefore holomorphic inside and on a simply closed contour ${\cal C}$ cutting the real axis at the points $-1/2$ and 
$m + 1/2$. As result thereof, the Euler Summation Formula
$$\sum\limits_{k=0}^m f(k) = \int\limits_0^m f(t)dt + (1/2)[f(0) + f(m)] - \int\limits_0^m 
\Omega_1(t)f^\prime(t)dt,\quad \Omega_1(t) = 2\sum\limits_{\nu=1}^\infty {sin (2\pi \nu t) \over 2\pi \nu}\eqno(2.11)$$
 holds\sref{\BS} on the one hand. On the other hand, we may also apply the concept of the total residue\sref{\BS}, in particular
$$ \sum\limits_{k=0}^m f(k) = {1\over 2\pi i} {\oaint}\ \pi\ ctg(\pi z) \cdot f(z)dz - \sum Res_{\cal C} \{\pi\ ctg(\pi z)\cdot  f(z)\},\eqno(2.12)$$
where $\sum Res_{\cal C}$  stands for the sum of the residues of the given function at all singular points inside the closed contour ${\cal C}$. If we let ${\cal C}_1$ and ${\cal C}_2$ be the arcs of contour ${\cal C}$ lying above and below the real axis respectively ($-1/2$ is the endpoint of 
${\cal C}_1$ and $m + 1/2$ that of ${\cal C}_2$), then\sref{\BS} 
$$\eqalign{ \sum\limits_{k=0}^m f(k) &=  \int\limits_{-1/2}^{m+1/2} f(t)dt - \int\limits_{{\cal C}_1} {f(z) \over e^{-2 \pi i z} - 1}dz + \int\limits_{{\cal C}_2} {f(z) \over e^{2 \pi i z} - 1}dz, \cr
&f(z) = e^{i \pi z}
{\Gamma(z + \gamma - \alpha /2)\Gamma(\alpha /2 - z + n)
\over \Gamma(z + 1)\Gamma(m + 1 - z) \Gamma(z + \gamma)\Gamma(\alpha /2 - z)}.}\eqno(2.13)$$
The two summation formulas (2.11) and (2.13) are already of great theoretical as well as practical value, although the function $f(z)$ of (2.10) is still somewhat cumbersome since $\Gamma (z)$ is determined in terms of improper integrals.  Fortunately we are able to derive closed-form formulas for these and also for our summation expression (2.6) by the use of the higher order hypergeometric function $_3 F_2$. Therewith we obtain the exact value of the summation (2.6). 

We further simplify the expression (2.9) by taking note of 
$$
(a-k)_n={(1-a)_k(a)_n\over (1-a-n)_k}
$$ 
in order to justify the relation
$$
\eqalign{x^{-\alpha}_{mn}&=
{1\over 2}T_nT_m 
B^{{\alpha\over 4}-{\gamma\over 2}}
{{\Gamma(\gamma-{\alpha\over 2})({\alpha\over 2})_n}
\over (\gamma)_n}
\sum\limits_{k=0}^m
{(-m)_k(\gamma-{\alpha\over 2})_k(1-{\alpha\over 2})_k
\over (\gamma)_k(1-{\alpha\over 2}-n)_k k!}\cr
&={1\over 2}T_nT_m 
B^{{\alpha\over 4}-{\gamma\over 2}}
{{\Gamma(\gamma-{\alpha\over 2})({\alpha\over 2})_n}
\over (\gamma)_n}
{}_3F_2(-m,\gamma-{\alpha\over 2},1-{\alpha\over 2};\gamma,1-{\alpha\over 2}-n
;1).}\eqno(2.14)
$$
After substituting for $T_n$ and $T_m$, we finally obtain
$$
x_{mn}^{-\alpha}=
(-1)^{m+n}
B^{\alpha\over 4}
\sqrt{{(\gamma)_n(\gamma)_m\over n! m!}}
{{\Gamma(\gamma-{\alpha\over 2})({\alpha\over 2})_n}
\over (\gamma)_n\Gamma(\gamma)}
{}_3F_2(-m,\gamma-{\alpha\over 2},1-{\alpha\over 2};\gamma,1-{\alpha\over 2}-n;1);\eqno(2.15)
$$
which is valid for all values of $\alpha$ and $\gamma$ such that $\gamma-{\alpha\over 2}\neq -k (k=0,1,\dots)$. Herein we treat the case $\alpha = 2$ by a limit, as follows: the limit of the hypergeometric function ${}_3F_2$ in Eq.(2.15) as $\alpha\rightarrow 2$ implies
$$
x_{mn}^{-2}=\cases{(-1)^{m+n}
{\sqrt{B}\over {\gamma-1}}
\sqrt{n!\over m!}{\sqrt{(\gamma)_n(\gamma)_m}\over (\gamma)_n}& if $n\geq m$,\cr
(-1)^{m+n}
{\sqrt{B}\over {\gamma-1}}
\sqrt{m!\over n!}{\sqrt{(\gamma)_m(\gamma)_n}\over (\gamma)_m}& if $m\geq n$.\cr}
\eqno(2.16)
$$

Two important observations follow from these results. First, the matrix elements of the Hamiltonian (1.2) are given by ($m,n=0,1,2,..,N-1$)
$$\eqalign{
H_{mn}=<m|H|n>\equiv 2\sqrt{B}(2n+\gamma)&\delta_{nm}+\lambda(-1)^{m+n}
B^{\alpha\over 4}
\sqrt{{(\gamma)_n(\gamma)_m\over n! m!}}
{{\Gamma(\gamma-{\alpha\over 2})({\alpha\over 2})_n}
\over (\gamma)_n\Gamma(\gamma)}\cr
&{}_3F_2(-m,\gamma-{\alpha\over 2},1-{\alpha\over 2};\gamma,1-{\alpha\over 2}-n;1)}\eqno(2.17)
$$
taken over the $N$-dimensional subspace spanned by basis (2.2). This is highly suitable for systematic calculations of the energy eigenvalues through diagonalization of the matrix 
$$\pmatrix{H_{00}&H_{01}&\dots&H_{0N-1}\cr
		    H_{10}&H_{11}&\dots&H_{1N-1}\cr
		    \dots&\dots&\dots&\dots\cr
		    H_{N-10}&H_{N-11}&\dots&H_{N-1N-1}}\eqno(2.18)$$  
with the aid of a computer.  By increasing the matrix dimension $N$, we can always improve these upper energy bounds. 

Secondly, the simple poles occurring at $\gamma-{\alpha\over 2}= 0$ can be removed for certain values of the parameter $\lambda$, if $\lambda=\gamma-{\alpha\over 2}$; thus the matrix elements (2.13) turn out to be
$$\eqalign{
H_{mn}\equiv 2\sqrt{B}(2n+\gamma)\delta_{nm}+(-1)^{m+n}
B^{\alpha\over 4}&
\sqrt{{(\gamma)_n(\gamma)_m\over n! m!}}
{{\Gamma(\gamma-{\alpha\over 2}+1)({\alpha\over 2})_n}
\over (\gamma)_n\Gamma(\gamma)}\cr
&{}_3F_2(-m,\gamma-{\alpha\over 2},1-{\alpha\over 2};\gamma,1-{\alpha\over 2}-n;1).}\eqno(2.19)
$$
These reduce, as $\lambda=\gamma-{\alpha\over 2}$ tends to zero, to
$$H_{mn}\equiv 2\sqrt{B}(2n+\gamma)\delta_{nm}+(-1)^{m+n}
{B^{\gamma\over 2}\over \Gamma(\gamma)}
\sqrt{{(\gamma)_n(\gamma)_m\over n! m!}}.\eqno(2.20)
$$
These expressions therefore, measure the vestigial effects of the remaining interactions\sref{\eks} after the perturbation $\lambda x^{-\alpha}$ is turned off along the path $\lambda=\gamma-{\alpha\over 2}\rightarrow 0$. We thus can identify certain paths for a given $\alpha$, where the matrix elements (2.13) always collapse to $2\sqrt{B}(2n+\gamma)\delta_{nm}$ as $\lambda\rightarrow 0$. Indeed, if we take for example $\sqrt{\lambda}=\gamma-{\alpha\over 2}$, then $A={1\over 4}(2\sqrt{\lambda}+\alpha-2)^2-{1\over 4}$ guarantees the existence of the matrix elements (2.13) for the $\alpha$ given; and further, as $\lambda$ tends to zero, the expressions (2.13) approach $2\sqrt{B}(2n+\gamma)\delta_{nm},$ as desired. There is nothing special about taking the square root of $\lambda$ of course, and it is certainly possible to identify other paths by selecting different powers of $\lambda$ for the given $\alpha$. This comment illustrates the fact, that suitable values of the parameter $A$ can lead to complete elimination of the vestigial effects of the potential as $\lambda\downarrow 0$.
\medskip

\noindent{\bf 3. Perturbation expansion for $\alpha<\gamma+1$}\medskip
\medskip
\noindent{\bf 3.1. The Eigenvalue Expansion}\medskip
A weak coupling perturbation expansion for a class of spiked harmonic oscillators, defined by (1.1) with zero angular momentum, was studied by Auguilera-Navarro {\it et al}\sref{\agub}. Their results have been expanded and further improved upon by the present authors\sref{\halb} in order to study the Hamiltonian (1.2). For sake of clarity, we shall summarize our results here. The standard perturbation theory to the second order, of the ground state energy of the generalized spiked harmonic oscillator Hamiltonian (1.2), yields the following expansion 
$$
E=E_0 +\lambda <0|x^{-\alpha}|0>+\lambda^2\sum\limits_{n=1}^\infty
{{|<0|x^{-\alpha}|n>|^2}\over E_0-E_n}+\dots,\eqno(3.1.1)
$$
where the $E_n, \ n=0,1,2,\dots,$ are the eigenenergies of the unperturbed Hamiltonian $H_0$ given by (2.3) and $<0|x^{-\alpha}|n>$ are expressed by means of (2.15) for $n=0,1,2\dots$. We have demonstrated that the weak-coupling expansion of the ground state energy of the Hamiltonian (1.2) leads to 
$$
\eqalign{
E=2\sqrt{B}\gamma +\lambda B^{\alpha\over 4} {\Gamma(\gamma-
{\alpha\over 2})\over \Gamma(\gamma)}&-\lambda^2 {B^{{\alpha-1}\over 2}
\alpha^2\over 16\gamma}{\Gamma^2(\gamma-{\alpha\over 2})\over \Gamma^2(\gamma)}
\cr
&
\times {}_4F_3(1,1,{\alpha\over 2}+1,{\alpha\over 2}+1;\gamma+1,2,2;1)+
\dots,} \eqno(3.1.2)
$$
wherein the generalized hypergeometric function ${}_4F_3$ is defined by (1.6). It is clear from the convergence condition of  
$
{}_4F_3
$ 
in Eq.(3.1.2), that regular perturbation fails in the case of $\alpha \geq \gamma+1$. It is also of interest to observe, that the series ${}_4F_3(1,1,{\alpha\over 2}+1,{\alpha\over 2}+1;\gamma+1,2,2;1)$ diverges as $\gamma-{\alpha\over 2}\rightarrow 0$.This divergence still holds, even if $\lambda=\gamma-{\alpha\over 2}\rightarrow 0,$ because in this case ${}_4F_3$ becomes ${}_3F_2(1,1,\gamma+1;2,2;1)$, which diverges for all $A\geq 0$. More precisely, the hypergeometric series ${}_3F_2(1,1,\gamma+1;2,2;1)$ diverge for all $A\geq -{1\over 4}$.

In the case of $\alpha=2$, ${}_4F_3(1,1,{\alpha\over 2}+1,{\alpha\over 2}+1;\gamma+1,2,2;1)$ reduced to ${}_2F_1(1,1;\gamma+1;1)$, which possesses a closed form sum given by ${\gamma\over \gamma-1}$. Thus the weak-coupling expansion for this case becomes
$$
E(\alpha=2)=2\sqrt{B}\gamma+{\sqrt{B}\over \gamma-1}\lambda-{\sqrt{B}\over 
4(\gamma-1)^3}\lambda^2+\dots,\eqno(3.1.3)
$$
which result coincides with the series expansion of $E=\sqrt{B}(2+\sqrt{1+4(A+\lambda)})$ about the point $\lambda=0$.

\medskip
\noindent{\bf 3.2. The Eigenfunction Expansion}\medskip
Several interesting questions have been raised by Aguilera-Navarro {\it et al} concerning the abnormal behavior of the standard weak coupling perturbation expansion in connection with attempts to derive the first-order correction to the wave function. Indeed, the first correction to the wave function by means of standard perturbation techniques, leads to
$$\psi_0^{(1)}(x)=\sum\limits_{n=1}^{\infty}{V_{n0}\over {E_0-E_n}}\psi_n(x),\eqno(3.2.1)$$
where $V_{n0}$ and $\psi_n(x)$ are given by Eq.(2.11) and (2.2) respectively. We have in this case
$$\psi_0^{(1)}(x)=-{B^{{\alpha\over 2}-{\gamma\over 4}-{1\over 2}}
\over 2\sqrt{2}}
{\Gamma(\gamma-{\alpha\over 2})\over \Gamma(\gamma)\sqrt{\Gamma(\gamma)}}
x^{\gamma-{1\over 2}}e^{-{\sqrt{B}\over 2}x^2}\sum\limits_{n=1}^\infty{({\alpha\over 2})_n\over n}{1\over n!}{}_1F_1(-n,\gamma,\sqrt{B}x^2).\eqno(3.2.2)
$$
To obtain a closed form for the sum appearing in Eq.(3.2.2) for $\alpha=2$, we use the relation between the confluent hypergeometric function ${}_1F_1(-n;\gamma+1;\cdot)$ and the associated Laguerre polynomials $L_n^{(\gamma)}(\cdot)$ given by\sref{\sla} 
$$
{}_1F_1(-n,\gamma+1,\cdot)={\Gamma(n+1)\Gamma(\gamma+1)\over \Gamma(n+\gamma+1)}
L_n^{(\gamma)}(\cdot).$$ Thus, the sum in Eq.(3.2.2) becomes
$$
\eqalign{
\sum\limits_{n=1}^\infty
{({1})_n\over n}{1\over n!}{}_1F_1(-n,\gamma,\sqrt{B}x^2)&=\Gamma(\gamma)
\sum\limits_{n=1}^\infty {(n-1)!\over \Gamma(n+\gamma)}L_n^{(\gamma-1)}(\sqrt{B}x^2)\cr
&=\Gamma(\gamma)[\psi(\gamma)-ln(\sqrt{B}x^2)],}\eqno(3.2.3)
$$
where we invoke formula 17 of the Bateman Project\sref{\Erd}, page 214, which originally was given by Toscano\sref{\ndlt}. Consequently, the first correction to the wave function, for the case of $\alpha=2$, takes the form $$ \psi_0^{(1)}(x)={B^{-{\gamma\over 4}} \over 2\sqrt{2}} {\Gamma(\gamma-1)\over \sqrt{\Gamma(\gamma)}} x^{\gamma-{1\over 2}}[ln(\sqrt{B}x^2)-\psi(\gamma)]e^{-{\sqrt{B}\over 2}x^2} . \eqno(3.2.4) $$ In the case of $\gamma=3/2$, or $A=0$, and $B=1$, Eq.(3.2.4) reads $$ \psi_0^{(1)}(x)= \pi^{-{1\over 4}} x [ln(x)-{1\over 2}\psi({3\over 2})]e^{-x^2/2}. $$ At this point we note that the coefficient of the exponential term is a logarithmic expression in variable $x,$ instead of being a polynomial.  This fact explains the odd result that the perturbation expansion fails to be regular. The difficulties of summing the expression that were reported by Aguilera-Navarro {\it et al}\sref{\agub}(\S III \& \S VI) arose from the existence of the alternating coefficient $(-1)^n$ in the matrix elements $V_{0n}$. However, we were able to overcome these difficulties by introducing the alternating coefficient $(-1)^n$ in the wave function $\psi_n(x)$.  It is an important result of the present work, that this is justified by the smooth transition of the exact solution of the Hamiltonian (2.1) to the exact solution of the known harmonic oscillator problem $(A=0)$ through the relation (2.4).

We now turn our attention to the evaluation of the sum in Eq.(3.2.2) for $\alpha <2$, given that $\gamma\geq {3\over 2}$ and $B$ is an arbitrary positive number. This requires a suitable integral representation of the confluent hypergeometric function ${}_1F_1(-n,\gamma,\sqrt{B} x^2)$ over an appropriate contour, in order to interchange summation with integration and thereby readily conclude the absolute convergence of the series just mentioned. We find the inverse Laplace transform (integral) representation $${}_1F_1(a,\gamma,\sqrt{B} z)=B^{{1\over 2}(1-\gamma)}{\Gamma(\gamma)\over 2\pi i}\int\limits_{c-i\infty}^{c+i\infty}e^{\sqrt{B} t}t^{-\gamma}(1-{z\over t})^{-a}dt\eqno(3.2.5)$$ under the conditions $\sqrt{B}$ is nonzero real, $Re(\gamma)>0, c>0,|arg(1-{z\over c})|<\pi$, with the replacement $z=x^2$ ($x$ real) to be most advantageous for achieving this end. Further, we choose $c$ sufficiently large to guarantee that $|1-{z\over t}|<\pi$ and also $|arg(1-{z\over c})|<\pi$ for $z=x^2$. Because we integrate over a path beginning and terminating with $c-i\infty$ and $c+i\infty$ respectively for variable $t=c+iy$, we must note  $$1-{z\over t}=1-{x^2(x-iy)\over c^2+y^2}=1-{x^2c\over c^2+y^2}+i{x^2y\over c^2+y^2}\quad (z=x^2),$$ for which $$|1-{z\over t}|^2=1-{x^2(2c-x^2)\over c^2+y^2}<1,$$ provided $c$ is choosen large enough - i. e. $x^2<2c$. Moreover, the condition $|arg(1-{z\over c})|<\pi$ must also be satisfied, which translates into $|arg(1-{x^2\over c})|<\pi$, and this is clearly true, provided $x^2<c$. Therefore, $c$ must be choosen so large that $c > x^2$. For such $c$ we shall always have $$0<1-{x^2(2c-x^2)\over c^2+y^2}<1\quad\quad \forall y\in R.$$  

Now we further continue the process of evaluating the summation in terms of the preceeding inverse Laplace-Transform representation (3.2.5) written for $a=-n$ and $z=x^2$, namely
$${}_1F_1(-n,\gamma,\sqrt{B} x^2)=B^{{1\over 2}(1-\gamma)}{\Gamma(\gamma)\over 2\pi i}\int\limits_{c-i\infty}^{c+\infty}e^{\sqrt{B} t}t^{-\gamma}(1-{x^2\over t})^n dt\quad (n=0,1,2,\dots),\eqno(3.2.6)
$$ which subsituted into the summation of Eq.(3.2.2) yields
$$\eqalign{ \sum\limits_{n=1}^\infty{({\alpha\over 2})_n\over n}&{1\over n!}{}_1F_1(-n,\gamma,\sqrt{B}x^2)=B^{{1\over 2}(1-\gamma)}(2\pi i)^{-1} \Gamma(\gamma)\sum\limits_{n=1}^\infty {({\alpha\over 2})_n\over n}{1\over n!}\int\limits_{c-i\infty}^{c+i\infty}e^{\sqrt{B} t}t^{-\gamma}(1-{x^2\over t})^n dt\cr &=B^{{1\over 2}(1-\gamma)}(2\pi)^{-1} \Gamma(\gamma)\sum\limits_{n=1}^\infty {({\alpha\over 2})_n\over n}{1\over n!} \int\limits_{-\infty}^{\infty}e^{\sqrt{B} (c+iy)}(c+iy)^{-\gamma} (1-{x^2\over c+iy})^n dy.}\eqno(3.2.7)$$  
\nl The evaluation of this last infinite sum, involving integrations over the interval  ($-\infty,\infty$), is achieved by examining the summation of the integrands, namely
$$ \sum\limits_{n=1}^\infty {({\alpha\over 2})_n\over n}{1\over n!} e^{\sqrt{B} (c+iy)}(c+iy)^{-\gamma}(1-{x^2\over c+iy})^n =e^{\sqrt{B} (c+iy)}(c+iy)^{-\gamma}\sum\limits_{n=1}^\infty {({\alpha\over 2})_n\over n}{1\over n!}(1-{x^2\over c+iy})^n,\eqno(3.2.8) $$
and demonstrating that it has an $L_1(-\infty,\infty)$-majorant. Hence, the existence of such a majorant shall permit us to interchange summation with integration, as result of the Lebesgue Dominated Convergence Theorem\sref\jowe. To arrive at such a majorant, we continue by noting that $$ {(\alpha)_n\over n\ n!} = {\Gamma({\alpha\over 2}+n)\over n\Gamma({\alpha\over 2})\Gamma(n+1)}\approx {{ ({\alpha\over 2}+n)^{{\alpha\over 2}+n-{1\over 2}} e^{-{\alpha\over 2}-n}\sqrt{2\pi}(1+o(1))}\over {n\Gamma({\alpha\over 2})(n+1)^{n+{1\over 2}}e^{-n-1}\sqrt{2\pi}(1+o(1))}}\eqno(3.2.9)$$
as consequence of the Stirling formula for large arguments of the gamma function. By means of  cancelations as well as utilizing the fact that
$${1+o(1)\over 1+o(1)}=1+o(1),$$
we may easily conclude that $${({\alpha\over 2})_n\over n\ n!}=  {{ ({\alpha\over 2}+n)^{{\alpha-1\over 2}+n}e^{1-{\alpha\over 2}}}\over \Gamma({\alpha\over 2})(1+n)^{{3\over 2}+n}}(1+o(1))= {(1+n)^{{\alpha\over 2}-2}\over \Gamma({\alpha\over 2})}(1+o(1)).\eqno(3.2.10) $$
Herein we utilize the well known sequence $(1 + {\rho\over n})^n$ defining the $e^\rho$, expressed asymptotically as  $$\bigg({{\alpha\over 2}+n\over 1+n}\bigg)^n=\bigg(1+ {{{\alpha\over 2}-1\over 1+n}}\bigg)^n = e^{{\alpha\over 2}-1}(1+o(1))\quad as\quad n\rightarrow \infty.\eqno(3.2.11)$$ We therefore have that $${{({\alpha\over 2})_n}\over n\ n!}={1\over \Gamma({\alpha\over 2})(n+1)^{2-{\alpha\over 2}}}[1+o(1)],\eqno(3.2.12)$$ which makes possible a comparison to the standard $p$-series $\sum\limits_{n=1}^{\infty}{1\over n^p}<\infty$ provided $p>1$. We thus have that $$\sum\limits_{n=1}^\infty {({\alpha\over 2})_n\over n\ n!}\leq A(\alpha)\sum\limits_{n=1}^\infty{1\over (n+1)^{2-{\alpha\over 2}}}<\infty \ {\sl with }\ A(\alpha)< \infty  \eqno(3.2.13)$$  for all $\alpha$ such that $2-{\alpha\over 2}>1$ or $\alpha<2$.   We now return to the majorization of summmation (3.2.8) in terms of inequality (3.2.13), which entails $$\eqalign{ \sum\limits_{n=1}^\infty { {({\alpha\over 2})_n}\over n\ n!}e^{\sqrt{B}(c+iy)}(c+iy)^{-\gamma}(1-{x^2\over c+iy})^n &\leq A(\alpha)\sum\limits_{n=1}^\infty (n+1)^{{\alpha\over 2}-2}e^{ c\sqrt{B}}|c+iy|^{-\gamma}|1-{x^2\over c+iy}|^n\cr &<A(\alpha,\sqrt{B},c)|c+iy|^{-\gamma}.}\eqno(3.2.14) $$
 Therein the constant $A(\alpha,\sqrt{B},c)$ is defined as $$ A(\alpha,\sqrt{B},c)=A(\alpha)e^{ c\sqrt{B}}\bigg[\sum\limits_{n=1}^\infty (n+1)^{{\alpha\over 2}-2}\bigg]\eqno(3.2.15) $$ and also $|1-{x^2\over c+iy}|<1$ was made use of. The most important aspect of inequality (3.2.14) is the appearance of the $L_1(-\infty,\infty)$-function $|c+iy|^{-\gamma}$ of variable $y$ majorizing the series $$\sum\limits_{n=1}^\infty { {({\alpha\over 2})_n}\over n\ n!}|e^{\sqrt{B}(c+iy)}(c+iy)^{-\gamma}(1-{x^2\over c+iy})^n|, $$ and this aspect justifies the evaluation of summation (3.2.7) by means of the Lebesgue Dominated Convergence Theorem\sref\jowe. Thus we specifically have
$$
\eqalign{ \sum\limits_{n=1}^\infty&{({\alpha\over 2})_n\over n}{1\over n!}{}_1F_1(-n,\gamma,\sqrt{B}x^2) = B^{{1\over 2}(1-\gamma)}{\Gamma(\gamma)\over  2\pi i} \sum\limits_{n=1}^\infty {({\alpha\over 2})_n\over n}{1\over n!} \int\limits_{-\infty}^{\infty}e^{\sqrt{B} (c+iy)}(c+iy)^{-\gamma}\times \cr  &(1-{x^2\over c+iy})^n idy =B^{{1\over 2}(1-\gamma)}{\Gamma(\gamma)\over 2\pi i} \int\limits_{-\infty}^{\infty}e^{\sqrt{B} (c+iy)}(c+iy)^{-\gamma} \bigg[\sum\limits_{n=1}^\infty {({\alpha\over 2})_n\over n}{1\over n!}(1-{x^2\over c+iy})^n\bigg]idy=\cr &B^{{1\over 2}(1-\gamma)}{\Gamma(\gamma)\over 2\pi} {\alpha\over 2} \int\limits_{-\infty}^{\infty}e^{\sqrt{B} (c+iy)}(c+iy)^{-\gamma} (1-{x^2\over c+iy}){}_3F_2(1,1,1+{\alpha\over 2};2,2;1-{x^2\over c+iy})dy,\ (3.2.16)} $$
which is an effective, straight-forward and precise determination of the summation $\sum\limits_{n=1}^\infty{({\alpha\over 2})_n\over n}{1\over n!}{}_1F_1(-n,\gamma,\sqrt{B}x^2)$ in terms of integrals of higher order hypergeometric function. However, by utilizing $t=c+iy$ we reconvert the last expression of relation (3.2.16) to the inverse Laplace transform format, namely  $$\eqalign{ \sum\limits_{n=1}^\infty&{({\alpha\over 2})_n\over n}{1\over n!}{}_1F_1(-n,\gamma,\sqrt{B}x^2)= \cr &B^{{1\over 2}(1-\gamma)}{\Gamma(\gamma)\over 2\pi i} {\alpha\over 2} \int\limits_{c-i\infty}^{c+i\infty}e^{\sqrt{B} t}t^{-\gamma} (1-{x^2\over t}){}_3F_2(1,1,1+{\alpha\over 2};2,2;1-{x^2\over t})dt.}\eqno(3.2.17)$$
The particality of this result lies in the fact that we can sum our series (3.2.7) involving hypergeometric function ${}_1F_1$ on any finite interval $(-c,c)$ in terms of either the Fourier Integral (3.2.16) or (even better) the inverse Laplace-Transform (3.2.17) involving the higher order hypergeometric function ${}_3F_2$ as given by the last integral expressions (3.2.16) and (3.2.17) respectively. This is exceedingly more effective than bringing in the Euler Summation Formula or the summation formula in terms of the total residues of $\pi[ctg(\pi z)]f(z)$.\bigskip
 
\noindent{\bf 4. Conclusion}\medskip
In this paper we have continued our study of the spiked harmonic-oscillator problem by expressing the Hamiltonian $H$ as a perturbation of the singular Gol'dman and Krivchenkov Hamiltonian $H_0.$ The zeroth-order eigenfunctions generated by $H_0$ form a suitable singularity-adapted basis for the appropriate Hilbert space.  Our principal results are threefold: (1) the derivation of the compact closed form (1.3) for the matrix elements $x^{-\alpha}_{mn};$ (2) the addition of a new expansion (1.4) for the wave function; and (3) the proof in \S (3.2) that the series for the wave function is convergent for all $\alpha \le 2.$  The compact matrix-element formula will certainly be useful for the generation of explicit eigenvalue estimates in particular cases.  Aguillera-Navaro and Guardiola\sref{\agub} have discussed the difficulties inherent in deriving a perturbation series for the wave function and they have presented a wish list of results to be sought.  We are happy to report that in the present work we have now found most of the results asked for in this list.
\bigskip
\noindent{\bf Acknowledgment}
\medskip Partial financial support of this work under Grant No. GP3438
from the 
Natural Sciences and Engineering Research Council of Canada is gratefully 
acknowledged.
\np
\references{1}
\bigskip

\end